\documentclass[12pt,a4]{article}
\usepackage{graphicx}
\usepackage{longtable}
\usepackage{amsmath}
\usepackage[all]{xy}

\newcommand{\be}{\begin{equation}}
\newcommand{\ee}{\end{equation}}
\newcommand{\ba}{\begin{eqnarray}}
\newcommand{\ea}{\end{eqnarray}}
\newcommand{\baa}{\begin{eqnarray*}}
\newcommand{\eaa}{\end{eqnarray*}}

\renewcommand{\k}{{\vec k}}

\def \r{\vec r}

\newcommand{\q}{{\vec q}}

\def \a{\vec a}
\def \K{{\vec K}}

\begin{document}

\title{The local density of states in the presence of impurity scattering in graphene at high magnetic field}
\author{Cristina Bena\\
{\small \it Laboratoire de Physique des Solides},
\vspace{-.1in}\\{\small \it B\^at 510, 91405 Orsay CEDEX, France}}
\maketitle

\begin{abstract}
We study the Fourier transform of the local density of states (LDOS) in graphene in the presence of a single impurity at high magnetic field.  We find that the most pronounced features occur for energies of the STM tip matching the Landau level energies. The Fourier transform of the LDOS shows regions of high intensity centered around the center and the corners of the Brillouin zone (BZ). The radial intensity dependence of these features is determined by the form of the wavefunctions of the electrons in the quantum Hall regime. Moreover, some of these regions break rotational symmetry, and their angular dependence is determined by the chirality of the graphene electrons. For the zeroth Landau level, the ratio between the features at the corners and center of the BZ depends on the nature of the disorder: it goes to zero for potential disorder, and is finite for hopping disorder. We believe that a comparison between our analysis and experiments will help understand the form of the quasiparticle wavefunction, as well as the nature of disorder in graphene.
\end{abstract}

The Friedel oscillations in the local density of states (LDOS) in graphene in the presence of impurity scattering have recently given rise to a lot of theoretical and experimental interest \cite{falko,glazman,bk,benaprl,others,mallet,mallet2,crommie,simon}. The low-energy Friedel oscillations resulting from intra-nodal scattering\footnote{We remind the reader that intra-nodal scattering denotes the scattering of a quasiparticle by an impurity such that a small change in its momentum occurs (the quasiparticle does not hop from one Dirac point to another), while inter-nodal scattering denotes scattering  in which a large change of momentum occurs, such that the quasiparticle hops between two Dirac points. Both processes conserve the energy of the quasiparticle.} decay atypically as $1/r^2$ \cite{falko,glazman}, while the oscillations due to inter-nodal scattering decay as $1/r$ \cite{benaprl}, as expected for two-dimensional systems \cite{bk,steveold}. This anomalous decay, and in some cases the breaking of the rotational symmetry of the high intensity features in the Fourier transform scanning tunneling spectra (FTSTS), are manifestations of the chirality of the electrons in graphene. These features have already been observed experimentally for epitaxial graphene \cite{mallet}. It is also interesting to note that for exfoliated graphene, the Friedel oscillations corresponding to intra-nodal scattering seem to decay as $1/r$ \cite{crommie}, and not as $1/r^2$, which may mean that some internal symmetry-breaking mechanism is at work in this system.

It appears therefore that the LDOS in the presence of impurity scattering in graphene can retrieve important information about the physics of its quasiparticles, specifically about their wavefunction. Here we use this to study the wavefunctions of the graphene quasiparticles in the quantum Hall effect (QHE) regime. This issue is of particular importance for understanding why the QHE arises in exfoliated, but not in epitaxial graphene, despite the presence of Landau levels (LL's) in both their spectra.

We focus on the regime of strong magnetic field (of order $50$T), when the magnetic length is of the order of $20-30$ lattice constants, and when the FTSTS features are easiest to discern. For much larger magnetic fields, the corresponding LL energy is too big, and no longer permits approximating the quasiparticle dispersion as linear. Furthermore, at huge magnetic fields (of order of thousands of T), when the magnetic length and the lattice constant become of the same order of magnitude, the intra-nodal and inter-nodal scattering features overlap\footnote{One should note that the distance between these features is given roughly by the inverse of the lattice constant, while their size is given by the inverse of the magnetic length.}, which makes the FTSTS features harder to interpret.
On the other hand, magnetic fields that are much smaller ($\le$ 1T) will give rise to Fourier-space features that are very sharply peaked and harder to analyze due to precision limitations.

For a given value of the magnetic field we calculate the FTSTS spectra when the tip bias matches the energy of the Landau levels, focusing in particular on the zeroth and the second LL. For energies of the tip situated between two Landau levels, within the approximations we use (Landau levels constant throughout the sample, single-impurity scattering, energy conservation), the intensity of the FTSTS spectra is negligible. We observe that for the zeroth LL, only high-intensity regions corresponding to intra-nodal scattering and to scattering between equivalent nodes are present (at the center of the BZ and reciprocal lattice points respectively), while no features can be identified at the corners of the BZ corresponding to scattering between nonequivalent nodes.
The scattering features are rotationally symmetric and decay in a Gaussian manner, consistent with the wavefunction of the quasiparticles in the zeroth LL of graphene.

For higher Landau levels, our calculations reveal both intra-nodal and inter-nodal scattering features. The intra-nodal ones, as well as the ones corresponding to scattering between equivalent nodes are rotationally symmetric. However, the patterns corresponding to scattering between nonequivalent nodes are asymmetric. Their asymmetry is a consequence of the chirality of the graphene quasiparticles. The radial dependence of these features stems from the electronic wavefunction of the Landau levels\footnote{The intensity is proportional to an integral of a Gaussian and two Hermite polynomials which ends up proportional to a Laguerre polynomial.}, and shows intensity minima and maxima. The scale of these fluctuations, as well as the scale associated with the Gaussian decay, are proportional to the inverse magnetic length.

\bigskip

The tight-binding Hamiltonian for monolayer graphene is:
\be
{\cal H}=\int d^2 \vec{k} [a_{\vec{k}}^{\dagger} b_{\vec{k}} f(\vec{k})+h.c.],
\label{h0}
\ee
where the operators $a^{\dagger}$, $b^{\dagger}$ correspond to creating electrons on the sublattice
$A$ and $B$ respectively, and \be f(\k)= -t ( e^{i \k \cdot \a_1} +  e^{i \k \cdot \a_2} +1). \label{f1} \ee Here
$\vec{a}_1 \equiv a(\sqrt{3} \hat{x}+3 \hat{y}/2)$, $\vec{a}_2 \equiv a(-\sqrt{3} \hat{x}+3\hat{y}/)2$, $t$ is the nearest-neighbor hopping amplitude, and
$a$ is the spacing between two adjacent carbon atoms, which we are setting to $1$.

As well known, the energy vanishes at the Dirac points, which are at (see e.g. \cite{gilles})
$$\K_{\mu\nu}^{\xi}= {\xi} {\a_1^*-\a_2^* \over 3}+ \mu \a_1^* + \nu \a_2^*$$
Here $\xi=\pm$ is the valley index (there are two such nonequivalent points for each elementary cell of the reciprocal space), $a_1^*=(2 \pi / \sqrt{3} a, 2\pi / 3 a)$, and $a_2^*=(-2 \pi / \sqrt{3} a, 2\pi / 3 a)$, and ($\mu$,$\nu$) span the family of equivalent Dirac points in the reciprocal space.
Note that
$$\K_{\mu\nu}^{\xi} \cdot \a_1= {2 \pi \xi \over 3}+ 2 \mu \pi \, , \qquad
\K_{\mu\nu}^{\xi} \cdot \a_2= -{2 \pi \xi \over 3}+ 2 \nu \pi.$$
Thus the Hamiltonian can be expanded around the Dirac points $\k= \K^{\xi}_{\mu\nu} + \q$ to find\footnote{Note that here we use a different convention for the definition of the Fourier transform than in \cite{gilles} which yields an opposite sign for $f^{\xi}(k)$.}
\be f^{\xi}(\k)= -v (\xi q_x - i q_y)\ee where $v=3 t/(2 a)$.
We work with the linearized Hamiltonian, and take into account the quasiparticles associated with all the Dirac points in the system. This can be done by adding the new indices $\xi$, $\mu$, and $\nu$ to the wavefunction of the quasiparticles, which characterize the position of the corresponding Dirac point $\K_{\mu\nu}^{\xi}$.

In the absence of magnetic field, the eigenfunctions of the above Hamiltonian have been extensively studied (see e.g.\cite{review} and references therein). In the presence of a large magnetic field (QHE regime), the eigenfunctions have also been determined in Refs.~\cite{castroneto,brey} by noting that the Hamiltonian reduces to the Hamiltonian of the harmonic oscillator. The diagonalization of the Hamiltonian for $\xi=1$, $\mu=0$, and $\nu=0$ (around the point $\K_{00}^{1}$) can be done by building the eigenfunction \cite{castroneto,brey}:
\begin{align}
\Psi(\r)=\sum_k \frac{e^{i k x}}{\sqrt{L}}\begin{pmatrix}0\\\phi_0(y-k l_B^2)\end{pmatrix} c_{k,-1}+\sum_{k,n,\alpha}\frac{e^{i k x}}{\sqrt{2 L}}\begin{pmatrix}\phi_n(y-k l_B^2)\\ \alpha \phi_{n+1}(y-k l_B^2)\end{pmatrix}c_{k,n,\alpha}\,,
\end{align}
where $l_B=\sqrt{\hbar/e B}\approx 26nm/\sqrt{B[T]}$ is the magnetic length, and $\phi_n(y)(n=0,1,2...)=e^{-y^2/2 l_B^2} H_n(y)$ are the eigenfunctions of the one-dimensional harmonic oscillator ($H_n(y)$ are the usual Hermite polynomials). Also, $\r=(x,y)$, the $c_{k,n,\alpha}$'s are the annihilation operators for quasiparticles in the $n+1$'st LL, with wavenumber $k$ along the $x$ direction and band $\alpha$, and $c_{k,-1}$ is the annihilation operator for a quasiparticle in the zeroth LL. In the new ``$c$''-operator basis the Hamiltonian is diagonal, and the Green's functions are:

\be
G_{n,k,\alpha}(\omega)=\langle c^{\dagger}_{n, k, \alpha}(\omega) c_{n,k,\alpha}(\omega) \rangle=\frac{1}{ \omega+i \delta-E_{n,k,\alpha}}\,,
\ee

We generalize this eigenfunction to take into account all the Dirac points (for the first BZ this reduces to the wavefunctions described in \cite{brey}) and we obtain:
\be
\Psi(\r)=\sum_{\xi=\pm 1,\mu,\nu} \Psi^{\xi}_{\mu \nu}(\r) e^{i \K_{\mu \nu}^{\xi}\cdot \r}\,,
\label{psi}
\ee
with
\ba
\Psi^{\xi}_{\mu\nu}(\r)=&&\sum_k \frac{e^{i k x}}{2 \sqrt{L}}\begin{pmatrix}(1-\xi)\phi_0(y-k l_B^2)\nonumber
\\ (1+\xi) \phi_0(y-k l_B^2)\end{pmatrix} c_{k,-1,\xi}^{\mu \nu}\\&&+\sum_{n,k,\alpha}\frac{e^{i k x}}{2\sqrt{2 L}}\begin{pmatrix}(1+\xi)\phi_n(y-k l_B^2)-\alpha (1-\xi)\phi_{n+1}(y-k l_B^2)\\ \alpha (1+\xi) \phi_{n+1}(y-k l_B^2)+(1-\xi) \phi_{n}(y-k l_B^2)\end{pmatrix}c_{k,n,\alpha,\xi}^{\mu\nu}\,.
\ea
The $c_{k,n,\alpha,\xi}^{\mu\nu}$-operators are annihilation operators  that beside the wavenumber $k$, band $\alpha$ and LL index $n$ have also the valley indices $\xi$, $\mu$ and $\nu$. Their unperturbed correlation functions do not depend on the valley indices:
\be
G_{n,k,\alpha,\xi}^{\mu \nu}(\omega)=\langle c^{\dagger \mu \nu}_{n, k, \alpha,\xi}(\omega) c_{n,k,\alpha,\xi}^{\mu \nu} (\omega)\rangle=\frac{1}{\omega+i\delta-\alpha E_{n}}\,,
\label{gf}
\ee
and, in the absence of disorder, correlators of operators connecting two different valleys are zero.

We introduce a delta-function impurity localized on an atom belonging for example to the $A$ sublattice with an impurity potential
\be
V=u \Psi_A^{\dagger}(\r=0)\Psi_A(\r=0)\,,
\ee
where $\Psi(\r)$ is given by Eq.(\ref{psi}). Using Eqs.(\ref{psi},\ref{gf}) and the Born approximation, we find that in the presence of the impurity potential $V$, the corrections to the correlation functions for the $c$ operators are given by:
\ba
\delta \langle c_{k,-1,\xi}^{\dagger \mu \nu} c_{k',-1,\xi'}^{\mu' \nu'}\rangle&=&u \frac{1}{(\omega+i\delta)^2} (1-\xi)(1-\xi')\phi_0(-k l_B^2) \phi_0(-k' l_B^2)\nonumber \\
\delta \langle c_{k,-1,\xi}^{\dagger \mu \nu} c_{k',n,\alpha,\xi'}^{\mu' \nu'}\rangle&=&u \frac{1}{(\omega+i\delta)(\omega+i \delta-\alpha E_{n})}[ (1-\xi)(1+\xi')\phi_0(-k l_B^2) \phi_n(-k' l_B^2)\nonumber \\&&
-\alpha(1-\xi)(1-\xi') \phi_0(-k l_B^2)\phi_{n+1}(-k' l_B^2)]\nonumber \\
\delta \langle c_{k,n,\alpha,\xi}^{\dagger \mu \nu} c_{k',-1,\xi'}^{\mu' \nu'}\rangle&=&u \frac{1}{ (\omega+i\delta)( \omega+i\delta-\alpha E_{n})}[ (1+\xi)(1-\xi')\phi_n(-k l_B^2) \phi_0(-k' l_B^2)\nonumber \\&&
-\alpha(1-\xi)(1-\xi') \phi_{n+1}(-k l_B^2)\phi_{0}(-k' l_B^2)]\nonumber \\
\delta \langle c_{k,n,\alpha,\xi}^{\dagger \mu \nu} c_{k',n',\alpha',\xi'}^{\mu' \nu'}\rangle&=&u \frac{1}{(\omega+i\delta-\alpha E_{n})(\omega+i\delta-\alpha' E_{n'})}[ (1+\xi)(1+\xi')\phi_n(-k l_B^2) \phi_n'(-k' l_B^2)\nonumber \\&&-
\alpha'(1+\xi)(1-\xi')\phi_{n}(-k l_B^2) \phi_{n'+1}(-k' l_B^2)\nonumber \\&&-\alpha(1-\xi)(1+\xi')\phi_{n+1}(-k l_B^2) \phi_n'(-k' l_B^2)\nonumber \\&&+\alpha \alpha'(1-\xi)(1-\xi')\phi_{n+1}(-k l_B^2) \phi_{n'+1}(-k' l_B^2)]
\label{expr}
\ea

We can now use these formulas to compute the corrections to the LDOS due to impurity scattering. Given that the expectation value of the density operator is \cite{gilles,benaprb}:
\be
\rho_{A/B}(\r)=\langle \Psi_{A/B}^\dagger(\r) \Psi_{A/B}(\r) \rangle
\ee
with the $A$ and $B$ components of the $\Psi(\r)$ being given by Eq.(\ref{psi}), we can write the Fourier transform of the LDOS as:
\be
\rho(\q,\omega)=\int d\r e^{-i \q \cdot \r}[\rho_A(\r) +\rho_B(\r) e^{-i \q \cdot \delta_{AB}}]
\ee
(the inclusion of the phase factor in the second term is explained in Refs.\cite{gilles,benaprb}).
A long but straightforward calculation yields for $\delta \rho(\q)$ due to impurity scattering:
\ba
\delta \rho_0(\q)&\propto&\sum_{\xi,\xi',\mu,\nu,\mu',\nu'}\int_{-\infty}^\infty dk \int_{-\infty}^{\infty} dy e^{-i q_y y} e^{i(K_{\mu' \nu' y}^{\xi'}-K_{\mu \nu y}^\xi)y} (1-\xi)^2 (1-\xi')^2 \frac{1}{(\omega+i \delta)^2}\times
\nonumber \\&&\phi_0(y-k l_B^2) \phi_0(y-k' l_B^2) \phi_0(-k l_B^2) \phi_0(-k' l_ B^2)|_{k'=k-q_x+K_{\mu'\nu'x}^{\xi'}-K_{\mu\nu x}^{\xi}}
\label{zero}
\ea
for the zeroth LL ($\omega=0$), and
\ba
\delta \rho_{n \alpha}(\q)&\propto&\sum_{\xi,\xi',\mu,\nu,\mu',\nu'}\int_{-\infty}^\infty dk \int_{-\infty}^{\infty} dy e^{-i q_y y} e^{i(K_{\mu' \nu' y}^{\xi'}-K_{\mu \nu y}^\xi)y} \frac{1}{(\omega-\alpha E_{n}+i \delta)^2}\times
\nonumber \\&&\times \{\phi_n(y-k l_B^2) \phi_n(y-k' l_B^2)[(1+\xi)(1+\xi')+e^{-i q_y a} (1-\xi)(1-\xi')]\nonumber\\&&-\alpha \phi_n(y-k l_B^2) \phi_{n+1}(y-k' l_B^2)[(1+\xi)(1-\xi')-e^{-i q_y a} (1-\xi)(1+\xi')]
\nonumber\\&&-\alpha \phi_{n+1}(y-k l_B^2) \phi_{n}(y-k' l_B^2)[(1-\xi)(1+\xi')-e^{-i q_y a}(1+\xi)(1-\xi')]\nonumber\\&&+\phi_{n+1}(y-k l_B^2) \phi_{n+1}(y-k' l_B^2)[(1-\xi)(1-\xi')+e^{-i q_y a}(1+\xi)(1+\xi')]\}\times
\nonumber\\&&
\times [(1+\xi)(1+\xi')\phi_n(-k l_B^2) \phi_n(-k' l_B^2)-\alpha (1+\xi)(1-\xi')\phi_n(-k l_B^2) \phi_{n+1}(-k' l_B^2)\nonumber \\&&-\alpha (1-\xi)(1+\xi')\phi_{n+1}(-k l_B^2) \phi_n(-k' l_B^2)\nonumber \\&&+(1-\xi)(1-\xi')\phi_{n+1}(-k l_B^2) \phi_{n+1}(-k' l_B^2)]|_{k'=k-q_x+K_{\mu'\nu'x}^{\xi'}-K_{\mu\nu x}^{\xi}}\}
\ea
for the $n+1$'st LL and band $\alpha$ ($\omega=\alpha E_n$).

In this calculation we have neglected the contributions coming from a quasiparticle being scattered between different LL's, as we work under the simplifying assumptions that the dominant scattering mechanism is elastic, and that the energy of each LL is constant throughout the sample. We have also focused on energies matching the energies of the Landau levels. For intermediate energies the intensity of the spectra is greatly reduced in the limit $\delta \rightarrow 0$, as it can be seen from the factor of $1/[(\omega+i\delta-\alpha E_n)(\omega+i \delta -\alpha' E_{n'})]$ in Eq.~(\ref{expr}).

For the zeroth LL ($\omega=0$) we compute $\delta \rho(\q,\omega)$ analytically:

\be
\delta \rho_0(\q)=\sum_{\mu,\nu,\mu',\nu'}\delta \tilde{\rho}_0(\q+\K^{-1}_{\mu \nu}-\K^{-1}_{\mu'\nu'})\,,
\ee
where
\ba
\delta \tilde{\rho}_0(\q)&\propto& \int_{-\infty}^\infty dy e^{-i q_y y} e^{-(y/l_B)^2/2} e^{-(y/l_B+q_x l_b)^2/2}
\int_{-\infty}^\infty dk e^{i q_y k l_B^2} e^{-( k l_B)^2/2} e^{-(k l_B+q_x l_b)^2/2} \nonumber \\
&\propto&  e^{-l_B^2 (q_x^2+q_y^2)/2}\,.
\ea

\begin{figure}[htbp]
\begin{center}
\includegraphics[width=3.5in]{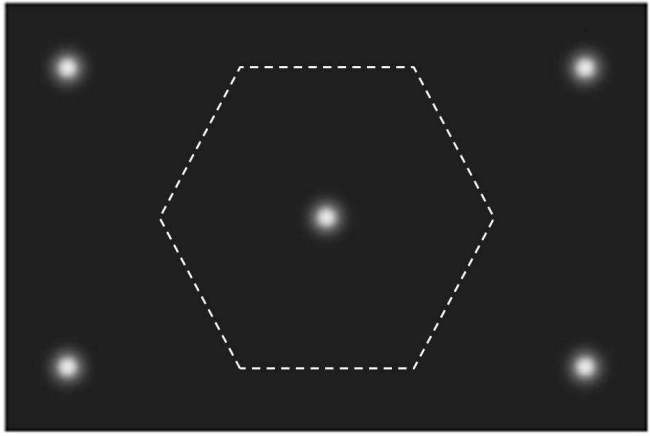}
\vspace{0.15in} \caption{\small FTSTS spectrum for a monolayer graphene sample with a single delta-function impurity, for an energy $E=0$ inside the zeroth LL, and $l_B/a=7$. The BZ is indicated by dashed lines.}
\label{fig1}
\end{center}
\end{figure}

The corresponding spectrum is plotted in Fig.~\ref{fig1}.
We note that there are regions of high intensity corresponding to quasiparticle scattering between two equivalent nodes (at the center of the first BZ and all equivalent points related via translation by a reciprocal lattice vector), but no high-intensity regions at the corners of the BZ corresponding to scattering between nonequivalent nodes. This can be seen directly from Eq.~(\ref{zero}) as only the term proportional to $(1-\xi)^2(1-\xi')^2$ appears in $\delta \tilde{\rho}(\q,\omega)$ in the zeroth LL, and this term is nonzero only if $\xi=\xi'=1$. This is related to the fact that in the zeroth LL the electronic wavefunctions have only one non-zero component (A or B), depending on the type of node ($\xi=\pm 1$) on which the electron sits. In order for the electron to scatter between two nodes, it needs to be able to change the sublattice index during the scattering process. However, for the type of impurities we consider (potential disorder), this is not possible. The shape of the observed regions is rotationally symmetric, and the intensity decays with the distance from the center in a manner characteristic to the decay of  the ground state of the harmonic oscillator ($e^{-x^2}$).

For an energy corresponding to the $n+1$'st LL, the FT of the LDOS is given by:

\be
\delta \rho_{\alpha n}(\q)=\sum_{\mu,\nu,\mu',\nu',\xi,\xi'}\delta\tilde{\rho}_{\alpha n \mu\nu\mu'\nu'}^{\xi \xi'}(\q+\K^{\xi}_{\mu \nu}-\K^{\xi'}_{\mu'\nu'})\,,
\ee
where
\ba
&\delta \tilde{\rho}_{\alpha n\mu \nu\mu'\nu'}^{\xi \xi'}(\q)&\propto \nonumber\\&&\int_{-\infty}^\infty dy e^{-i q_y y} \{\phi_n(y) \phi_n(y+q_x l_B^2)[(1+\xi)(1+\xi')+ e^{-2 i \pi(\mu+\nu-\mu'-\nu')/3}(1-\xi)(1-\xi')]\nonumber\\&&-\alpha \phi_n(y) \phi_{n+1}(y+q_x l_B^2)[(1+\xi)(1-\xi')- e^{-2 i \pi(\mu+\nu-\mu'-\nu')/3}(1-\xi)(1+\xi')]
\nonumber\\&&-\alpha \phi_{n+1}(y) \phi_{n}(y+q_xl_B^2)[(1-\xi)(1+\xi')-e^{-2 i \pi(\mu+\nu-\mu'-\nu')/3}(1+\xi)(1-\xi')]\nonumber\\&&+\phi_{n+1}(y) \phi_{n+1}(y+q_x l_B^2)[(1-\xi)(1-\xi')+e^{-2i \pi(\mu+\nu-\mu'-\nu')/3}(1+\xi)(1+\xi')]\}
\nonumber \\&& \times
\int_{-\infty}^\infty dk e^{-i q_y k l_b^2} [(1+\xi)(1+\xi')\phi_n(-k l_B^2) \phi_n(-k l_B^2+q_x l_B^2)\nonumber \\&&-\alpha (1+\xi)(1-\xi')\phi_n(-k l_B^2) \phi_{n+1}(-k l_B^2+q_x l_B^2)\nonumber \\&&-\alpha (1-\xi)(1+\xi')\phi_{n+1}(-k l_B^2) \phi_n(-k l_B^2+q_x l_B^2)\nonumber \\&&+(1-\xi)(1-\xi')\phi_{n+1}(-k l_B^2) \phi_{n+1}(-k l_B^2+q_x l_B^2)]\,,
\ea
and where we have considered that $q$ is small with respect to $1/a$, hence approximating $e^{-i q_y a}\approx 1$.
Denoting $I_{m n}(\q)\equiv \int_{-\infty}^{\infty} e^{-i q_y y} \phi_m(y) \phi_n(y+q_x l_B^2)$, we can rewrite the above formula as:
\ba
\delta \tilde{\rho}_{\alpha n\mu \nu\mu'\nu'}^{\xi \xi'}(\q)&\propto& \{(1+\xi)^2(1+\xi')^2 |I_{nn}(\q)|^2+(1-\xi)^2(1-\xi')^2 |I_{n+1,n+1}(\q)|^2\nonumber \\&&+(1+\xi)^2(1-\xi')^2|I_{n,n+1}(\q)|^2+(1-\xi)^2(1+\xi')^2|I_{n+1,n}(\q)|^2\nonumber \\&&
+e^{-2\pi i(\mu+\nu-\mu'-\nu')/3}[(1-\xi)^2(1-\xi')^2 I_{nn}(\q) I^*_{n+1,n+1}(\q)\nonumber \\&&
-(1-\xi)^2(1+\xi')^2 I_{n,n+1}(\q) I^*_{n+1,n}(\q)-(1+\xi)^2(1-\xi')^2 I_{n+1,n}(\q) I^*_{n,n+1}(\q)\nonumber \\&&+(1+\xi)^2(1+\xi')^2 I_{n+1,n+1}(\q) I^*_{n,n}(\q)]\}\,.
\label{rhoi}
\ea
We can compute the $I_{mn}$ integrals analytically:
\be
I_{mn}(\q)=\sqrt{\pi} 2^{N} M! [(\sigma_{nm}q_x l_B-i q_y l_B)/2]^{N-M} L_{M}^{N-M}(q^2l_B^2/2) e^{-(q_x-i q_y)^2 l_B^2/4}\,,
\label{imn}
\ee
where $L_m^n$ is a Laguerre polynomial, $\sigma_{nm}={\rm sign}(n-m)$, $q=|\q|$ is the length of the $\q$ vector, and $M$ and $N$ denote the smaller and respectively the larger of $m$ and $n$.

At this point there are a few observations we can make. The first is that $\delta \tilde{\rho}(\q,\omega)$ contains terms proportional to Laguerre polynomials (which in general characterize the overlap between two LL eigenfunctions\cite{rafael}). A Laguerre polynomial shows a number of zeroes (or ``nodes'') given by the order of the polynomial (e.g. $L_1$ has 1 node, $L_2$ has 2 nodes, and so on). As these Laguerrre polynomials appear in the dependence of the LDOS $\delta \rho(\q,\omega)$ on $q$, we expect $\delta \rho(\q,\omega)$ to also have a ``node'' structure. Indeed, the scattering between nonequivalent Dirac points ($\xi\ne\xi'$) gives rise to terms that contain only combinations of $I_{n,n+1}$ terms (all proportional to a Laguerre polynomial of order $n$); these combinations will have $n$ nodes. However, scattering between equivalent Dirac points $\xi=\xi'$  gives rise to terms that contain both $I_n$ and $I_{n+1}$, and hence contain combinations of Laguerre polynomials of different orders. In general one cannot predict the exact number of nodes of such combinations.

The second observation is that the high-intensity features show, besides the Laguerre polynomial dependence, a Gaussian decay on a scale of a few times the inverse magnetic length $e^{-q^2 l_B^2/4}$.

The third observation is that the expression in Eq.~(\ref{imn}) contains  rotationally asymmetric phase factors. The terms of the form $e^{i q_x q_y l_B^2/2}$ cancel in the final expression of $\tilde{\rho}$ where only products of an integral of the type $I_{mn}$ and of a complex conjugate of such an integral appear. The other complex phase factors $q_x\pm i q_y$ only appear if $m\ne n$, and hence, as it can be seen from Eq.~(\ref{rhoi}), only for processes involving scattering between nonequivalent nodes ($\xi \ne \xi'$). This type of scattering corresponds to coupling between the LL wavefunctions $\phi_n$ and $\phi_{n+1}$. For the non-relativistic QHE this coupling can only arise for the transition of a quasiparticle between two distinct (energy separated) Landau levels. However, we see that for graphene, due to the spinorial structure of the wavefunction, such ($\phi_n,\phi_{n+1}$) coupling occurs naturally inside the same LL, and appears to be a manifestation of the chirality of the graphene quasiparticles.

We have already discussed the particular case of the zeroth Landau level $n=-1$. To illustrate the observations above we focus on the FTSTS spectra in the second Landau level ($n=1$). This allows us to observe the node structure of the
results in more detail than for $n=0$ (first LL). The $I_{mn}$ integrals are given by:
\ba
I_{11}(\q)&=&\sqrt{\pi}(2-q^2 l_B^2) e^{-(q_x- i q_y)^2 l_B^2/4}\nonumber \\
I_{12}(\q)&=& \sqrt{\pi} (q_x-i q_y)(4-q^2 l_B^2) e^{-(q_x- i q_y)^2 l_B^2/4}  \nonumber \\
I_{21}(\q)&=&- \sqrt{\pi} (q_x+i q_y)(4-q^2 l_B^2)e^{-(q_x- i q_y)^2 l_B^2/4} \nonumber \\
I_{22}(\q)&=&  \sqrt{\pi} [8+q^2 l_B^2 (q^2 l_B^2-8)] e^{-(q_x- i q_y)^2 l_B^2/4}
\label{i}
\ea

\begin{figure}[htbp]
\begin{center}
\includegraphics[width=5in,angle=0]{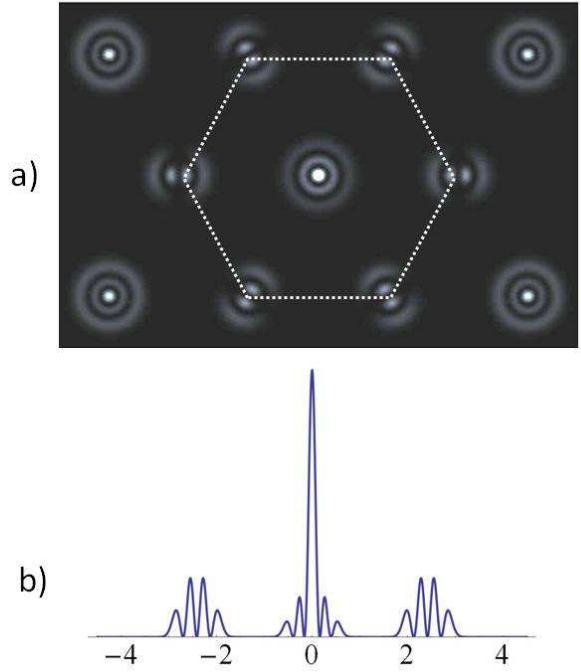}
\vspace{0.15in} \caption{\small a) FTSTS spectrum for a monolayer graphene sample with a single delta-function impurity, for an energy equal to the energy of the second LL i.e. $\omega=1.4V$, and for $l_B/a=7$, ($B=900T$).  The BZ is indicated by dashed lines. b) Horizontal cut through the spectrum depicted in a) for $k_y=0$ (in arbitrary units). For clarity, the intensity of the features at the corners of the BZ was multiplied by a factor of two.}
\label{fig3}
\end{center}
\end{figure}

In Figs.~2 and 3 we plot the real part of the FT of the LDOS, first for an unrealistically large magnetic field, $B=900T$ (Fig.~2) as well as for a physical $B=50T$ (Fig.~3). In the first case the corresponding energy for the second LL is of the order of $1.4V$ which makes the linear approximation for the spectrum invalid. However we use it to illustrate qualitatively our results, as the magnetic length is $l_B=7 a$, which gives rise to larger and better resolved features. For the more physical case of $B=50T$, as well as for smaller magnetic fields, these features should be observable for energies smaller or equal to $400meV$, for which the linear approximation is still reasonable \cite{linear}. However, at fields of the order of or smaller $50T$, the magnetic length is of the order of or larger than $26.5 a$, which gives rise to sharper features, therefore harder to resolve in momentum space, as it can be seen in Fig.~3.

\begin{figure}[htbp]
\begin{center}
\includegraphics[width=3.5in]{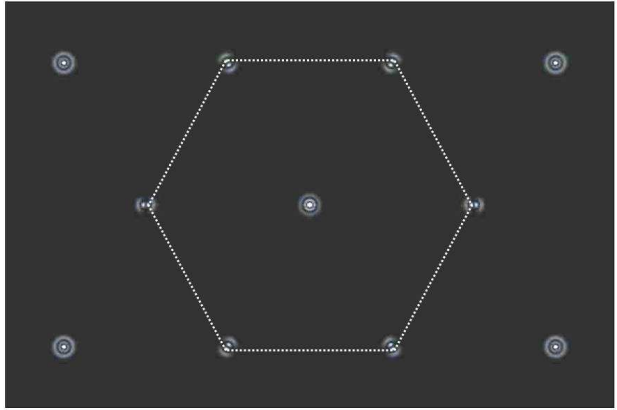}
\vspace{0.15in} \caption{\small The real part of the FTSTS spectra for a monolayer graphene sample with a single delta-function impurity, for an energy equal to the energy of the second LL i.e. $\omega \approx 400meV$, and $l_B/a=26.5$ ($B=50T$). The intensity of the features at the corners of the BZ was multiplied by a factor of two.}
\label{fig3}
\end{center}
\end{figure}

For both values of the magnetic field there exist regions of high intensity corresponding to both intra-nodal and inter-nodal scattering which exhibit maxima and minima superposed over a Gaussian decay; the distance between the maxima and minima is proportional to the inverse magnetic length. The high-intensity regions corresponding to scattering between nonequivalent Dirac points ($\xi \ne \xi'$) have only one zero-intensity node ($n=1$) in their radial dependence, as expected (see Fig.~\ref{fig3} a) and b)). The intensity also goes to zero at the center of these regions ($q=0$) because of the $q_x \pm i q_y$ factors in Eq.(\ref{i}). The high-intensity region corresponding to scattering inside the same Dirac point (located at the center of the BZ) has $n+1=2$ nodes, while the high-intensity features corresponding to scattering  between equivalent Dirac points ($\xi=\xi'$) show a structure of minima and maxima similar to that of the central feature, but no zero-intensity node.

Also, as noted above, the features corresponding to scattering between equivalent Dirac points are rotationally symmetric (depend only on the magnitude $q$). However, for scattering between nonequivalent Dirac points, $\rho(\q)$ is proportional to $(q_x\pm i q_y)^2$, which breaks rotational symmetry. This also happens in the imaginary part of the FTSTS spectrum, depicted in Fig.~4.

\begin{figure}[htbp]
\begin{center}
\includegraphics[width=3.5in,angle=180]{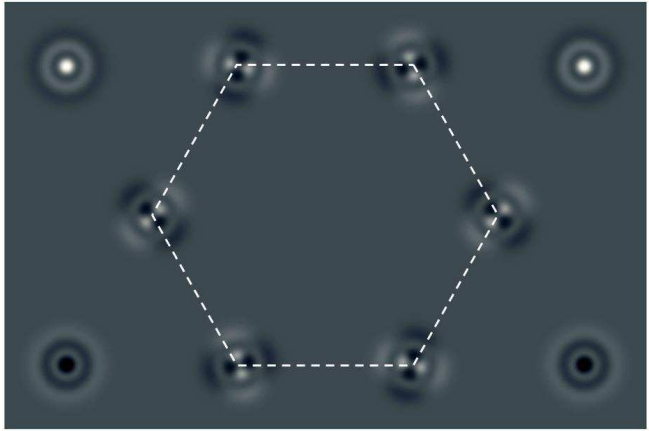}
\vspace{0.15in} \caption{\small The imaginary part of the FTSTS spectrum for a monolayer graphene sample with a single delta-function impurity, for  $\omega=1eV$, and $l_B/a=7$ ($B=900T$). As before, the intensity of the features at the corners of the BZ was multiplied by a factor of two.}
\label{fig3}
\end{center}
\end{figure}

\bigskip

To conclude, we have computed the effect of single-impurity scattering on the Fourier transform of the LDOS in the presence of a strong magnetic field. We have found that the FTSTS spectra contain high-intensity regions corresponding both to scattering processes in which quasiparticles remain at the same Dirac point (intra-nodal), and to scattering processes in which quasiparticles hop between different Dirac points (inter-nodal). Both types of processes give rise to features that contain information about the wavefunction of the electrons in the quantum Hall state (decay length, maxima, zeroes, minima). While scattering between equivalent nodes gives rise to rotationally symmetric features, the features coming from scattering between nonequivalent nodes break this symmetry, manifesting the chirality of the quasiparticles. A special situation arises for energies inside the zeroth LL when we observe no features corresponding to scattering between nonequivalent nodes. This should gives rise to a smoother spatial dependence of the LDOS than for the energies corresponding to higher Landau levels.

We focus on a localized impurity, but we expect our results to be quite similar for an extended Coulomb impurity. The most significant difference will be a reduction of the ratio between the intensity of the features away from the center and at the center. It would be interesting to see what happens if other types of disorder are considered which affect not only the electronic density, but also the hopping parameters in the neighborhood of the impurity. Unlike the potential disorder, the hopping disorder couples the nonequivalent valleys even in the zeroth LL, and one expects to observe the corresponding inter-nodal scattering features in the FTSTS spectra for all LLs. Therefore we propose to use the ratio between the intra-nodal and inter-nodal scattering features not only as a good indicator of the extension of the impurity potential, but also in the zeroth LL as an indicator of the form of the impurity potential (potential disorder versus hopping disorder).

The relevant physical regime for the magnetic field is between $20-50T$, and the dominant features we describe (for the second LL) are expected to arise for energies of the tip matching the energies of the LL's, i.e. $200-400meV$. It would be interesting to study what happens for energies of the tip situated between two Landau levels in the presence of multiple impurities. Because of disorder, the LL's are expected to broaden, which translates spatially into a LL energy that fluctuates spatially. Thus, for some energies, one may have overlapping contributions from various LL's, as well as more complicated (and not so neat) features arising in the FTSTS spectra.

We have found that the FTSTS spectra can give information about the electronic wavefunction in graphene at energies matching the LL energy (decay length, nodes, etc.), as well as about the nature of the disorder. We hope that a comparison between this theoretical study and experiments will shed light on the nature of the electronic states in epitaxial graphene under high magnetic field, where quantum-Hall features such as the LL's in the DOS are present, but no quantum Hall effect is observed. In particular it would be interesting to test whether the quasiparticles in epitaxial graphene are indeed described by typical quantum-Hall wavefunctions, and if so, whether it is possible to establish the nature of disorder in both epitaxial and in exfoliated graphene. Identifying the form of the wavefunction, as well as the nature of disorder will help understand why the quantum Hall effect is not observed in epitaxial graphene.

{\bf Acknowledgments} We would like to thank J.-N. Fuchs, M. Goerbig, G. Montambaux, and G. Rutter for useful discussions.

\end{document}